\documentclass
[showpacs,amsfonts,amssymb,oneside,notitlepage,10pt,a4paper,twocolumn,prl,preprint,nobalancelastpage]{revtex4}%
\usepackage{amsfonts}
\usepackage{amsmath}
\usepackage{amssymb}
\usepackage{graphicx}%
\providecommand{\U}[1]{\protect\rule{.1in}{.1in}}

\begin{document}
\title{Non-critically squeezed light via spontaneous rotational symmetry breaking}
\author{Carlos Navarrete--Benlloch, Eugenio Rold\'{a}n, and Germ\'{a}n J. de Valc\'{a}rcel}
\affiliation{Departament d'\`{O}ptica, Universitat de Val\`{e}ncia, Dr. Moliner 50,
46100--Burjassot, Spain.}

\begin{abstract}
We theoretically address squeezed light generation through the spontaneous
breaking of the rotational invariance occuring in a type I degenerate optical
parametric oscillator (DOPO) pumped above threshold. We show that a DOPO with
spherical mirrors, in which the signal and idler fields correspond to first
order Laguerre-Gauss modes, produces a perfectly squeezed vacuum with the
shape of a Hermite-Gauss mode, within the linearized theory. This occurs at
any pumping level above threshold, hence the phenomenon is non-critical.
Imperfections of the rotational symmetry, due e.g. to cavity anisotropy, are
shown to have a small impact, hence the result is not singular.

\end{abstract}

\pacs{42.50.L; 42.65.Sf}
\maketitle

\textit{Introduction.--} Squeezed light is a central tool in several
sophisticated applications of physics, high-precision measurements
\cite{Drummond04} and quantum information with continuous variables
\cite{Braunstein05} being probably the most outstanding. The quality of
squeezing, i.e. how less noisy light is as compared with vacuum (which sets
the so-called shot noise level) is a main concern for those applications as
any fluctuation level limits their performance. Improving the quality and
reliability of squeezing is thus an important goal.

The paradigmatic squeezing process is singlemode quadrature squeezing of an
optical field via degenerate parametric down conversion \cite{Yuen,Caves}, a
nonlinear process that converts a pump photon of frequency $2\omega_{0}$ into
two photons of frequency $\omega_{0}$. Although in this process perfect
squeezing is achieved only when the pump power goes to infinity, there is a
well known technique for increasing the squeezing level that consists in
confining the nonlinear interaction inside an optical cavity, in which case
one deals with a degenerate optical parametric oscillator (DOPO). In DOPOs
squeezing is ideally obtained at the oscillation threshold \cite{WM}: A
critical phenomenon. DOPOs are nowadays customarily utilized as sources of
squeezed light, reaching noise reductions as large as $10$dB below the shot
noise level (90\% of squeezing) \cite{9dB,10dB}.

Recently an alternative way for producing squeezed light was proposed by some
of us \cite{Perez}, based on the exploitation of the spontaneous translational
symmetry breaking occurring in a broad area, planar DOPO model. Such system
supports cavity solitons (CSs) --among other dissipative structures forming
across its transverse plane--, which are bright spots surrounded by darkness
that can be placed at any point in space, breaking the translational symmetry.
A study of their quantum fluctuations \cite{Perez} reveals that (i) the CS
position diffuses because of quantum noise, and (ii) a special transverse
mode, namely the $\frac{\pi}{2}$ phase shifted gradient of the CS (its linear
momentum), is perfectly squeezed at low fluctuation frequencies, irrespective
of the system's proximity to threshold. This is reminiscent of a Heisenberg
uncertainty relation, with the additional and compatible feature that the full
indetermination of the CS position in the long time limit is accompanied by
the perfect determination (perfect squeezing) of its momentum at low
frequencies, like a canonical pair in a minimum uncertainty state. A main
limitation of this result is that CSs have not been observed so far in DOPOs.
This is not a fundamental limitation but obviously reduces its practical
interest. Nevertheless it paves the way (non-critical squeezing via a
spontaneous spatial symmetry breaking) to other extensions, like the one we
consider here: The spontaneous rotational symmetry breaking of a DOPO, which
can be implemented with current technology. We hope that experiments based on
this new phenomenon will successfully generate high-quality non-critically
squeezed light.

\textit{Rotational invariance and squeezing: General description.--} Consider
a type I DOPO (i.e., with signal and idler photons degenerated both in
frequency and polarization) with spherical mirrors and pumped by a resonant,
coherent optical field of frequency $2\omega_{0}$ and Gaussian transverse
profile (i.e., pumped with zero orbital angular momentum photons). Such a
configuration is invariant under rotations around the cavity axis ($z$-axis).
The cavity is assumed to be tuned to the first transverse mode family at the
subharmonic frequency $\omega_{0}$, so the signal/idler field is a
superposition of two Laguerre-Gauss (L-G) modes, $L_{+1}\left(  \mathbf{r}%
\right)  $ and $L_{-1}\left(  \mathbf{r}\right)  $, which have opposite
orbital angular momenta. Any other transverse mode is assumed to be detuned
far enough from $\omega_{0}$. Inside this cavity a $\chi^{\left(  2\right)  }$
crystal down converts pump photons into signal/idler photon pairs, and vice
versa. These photons are degenerate in frequency because of energy
conservation and, because of orbital angular momentum conservation, each
photon pair must comprise one $L_{+1}$ photon plus one $L_{-1}$ photon. Hence
the number of $L_{+1}$ photons and $L_{-1}$ photons should be sensibly equal
and highly correlated. Another way of looking at this process follows from
noticing that the simultaneous emission of a $L_{+1}$ photon and a $L_{-1}$
photon corresponds to the emission of two photons in a first order
Hermite-Gauss (H-G), or TEM$_{10}$, mode, which breaks the rotational symmetry
of the system. The orientation of such mode in the transverse plane, measured
by the angle $\theta$ in Fig. 1, is determined by the relative phase between
the two subjacent L-G modes. The rotational invariance of the system implies
however that $\theta$ is arbitrary and one expects that quantum fluctuations
will induce a random rotation of the TEM$_{10}$ mode around the cavity axis.
In loose terms this means an \textquotedblleft
indefiniteness\textquotedblright\ in the value of $\theta$ that, generalizing
Ref. \cite{Perez}, should be acompanied by a reduction of fluctuations in the
canonically conjugated variable, the orbital angular momentum, associated to
the operator $-i\frac{\partial}{\partial\theta}$. We note that the angular
gradient of the TEM$_{10}$ mode is another H-G mode, spatially crossed with
respect to it, call it TEM$_{01}$ mode. Hence a balanced homodyne detection
that uses as a local oscillator a $\frac{\pi}{2}$ phase shifted TEM$_{01}$
mode should yield perfect squeezing at zero noise frequency at any pumping
level above threshold. This is the basic idea of squeezing generation via
symmetry breaking of the rotational invariance in DOPO, and below we
analytically demonstrate that this is what actually occurs \cite{nota0}.%
\begin{figure}
[h]
\begin{center}
\includegraphics[
height=5.329cm,
width=7.2994cm
]%
{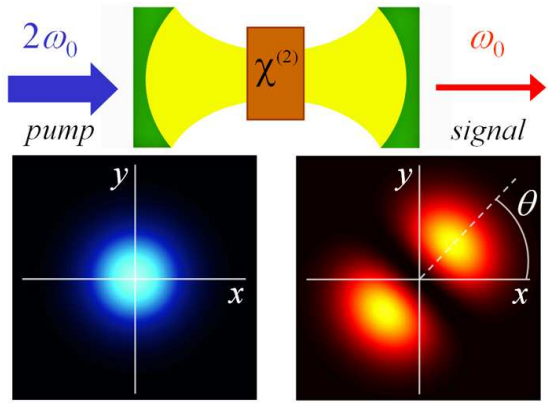}%
\caption{{\protect\small Scheme of the DOPO pumped by a Gaussian beam and
tuned to the first transverse mode family at the subharmonic. The orientation
}${\protect\small \theta}$ {\protect\small is arbitrary. }}%
\end{center}
\end{figure}

\textit{The model}.-- Inside the cavity there are three relevant modes: the
pumped Gaussian mode at frequency $2\omega_{0}$, and two L-G modes at the
subharmonic frequency $\omega_{0}$. The electric field at the cavity waist
(where the nonlinear crystal is assumed to be located) can be written as
\begin{equation}
\hat{E}\left(  \mathbf{r},t\right)  =i\mathcal{F}_{\mathrm{p}}\hat
{A}_{\mathrm{p}}\left(  \mathbf{r},t\right)  e^{-2i\omega_{0}t}+i\mathcal{F}%
_{\mathrm{s}}\hat{A}_{\mathrm{s}}\left(  \mathbf{r},t\right)  e^{-i\omega
_{0}t}+\mathrm{H.c.} \label{qfield}%
\end{equation}
where $\mathcal{F}_{\mathrm{p}}=\sqrt{\hbar\omega_{0}/n\varepsilon_{0}L}$,
$\mathcal{F}_{\mathrm{s}}=\mathcal{F}_{\mathrm{p}}/\sqrt{2}$, $L$ is the
effective cavity length, $n$ is the crystal refractive index,
\begin{subequations}
\label{SVE}%
\begin{align}
\hat{A}_{\mathrm{p}}\left(  \mathbf{r},t\right)   &  =\hat{a}_{0}\left(
t\right)  G\left(  \mathbf{r}\right)  ,\\
\hat{A}_{\mathrm{s}}\left(  \mathbf{r},t\right)   &  =\hat{a}_{+1}\left(
t\right)  L_{+1}\left(  \mathbf{r}\right)  +\hat{a}_{-1}\left(  t\right)
L_{-1}\left(  \mathbf{r}\right)  , \label{As}%
\end{align}
are slowly varying envelopes, and $\hat{a}_{m}\left(  t\right)  $ and $\hat
{a}_{m}^{\dagger}\left(  t\right)  $ are the interaction picture boson
operators for each mode ($m=0,\pm1$) obeying $\left[  \hat{a}_{m}\left(
t\right)  ,\hat{a}_{n}^{\dagger}\left(  t\right)  \right]  =\delta_{mn}$. The
Gauss, $G\left(  \mathbf{r}\right)  $, and L-G, $L_{\pm1}\left(
\mathbf{r}\right)  $, mode envelopes are given by \cite{Siegman} $G\left(
\mathbf{r}\right)  =\sqrt{2}\pi^{-1/2}w^{-1}e^{-r^{2}/w^{2}}$ and $L_{\pm
1}\left(  \mathbf{r}\right)  =\pi^{-1/2}w^{-2}re^{-r^{2}/2w^{2}}e^{\pm i\phi}%
$, $r$ and $\phi$ are the polar coordinates in the transverse plane, and $w$
($\sqrt{2}w$) is the beam radius of the pump (signal) beam at its waist.

The interaction Hamiltonian describing pumping and the nonlinear mixing
processes ocurring at the nonlinear crystal reads $\hat{H}=i\hbar\left(
\mathcal{E}_{\mathrm{p}}\hat{a}_{0}^{\dagger}+\chi\hat{a}_{+1}^{\dagger}%
\hat{a}_{-1}^{\dagger}\hat{a}_{0}\right)  +\mathrm{H.c.}$,\textrm{ }where
$\mathcal{E}_{\mathrm{p}}$ is the amplitude of the external coherent pump,
real without loss of generality, and $\chi$ is the nonlinear coupling constant
\cite{chi}. Assuming that pump and signal modes are damped at rates
$\gamma_{\mathrm{p}}$ and $\gamma_{\mathrm{s}}$, respectively, losses
occurring at only one cavity mirror (which needs not be the same for pump and
signal), one can write down Langevin equations using well known techniques of
quantum optics of open systems. We notice however that our Hamiltonian is
formally equivalent to that for the nondegenerate OPO and use the
corresponding Langevin equations in the positive $P$ representation as given
in \cite{Dechoum}. This representation sets a correspondence between operators
$\left\{  \hat{a}_{m}\left(  t\right)  ,\hat{a}_{m}^{\dagger}\left(  t\right)
\right\}  $ and independent c-number stochastic variables $\left\{  \alpha
_{m}\left(  t\right)  ,\alpha_{m}^{+}\left(  t\right)  \right\}  $
respectively, so that any stochastic average equals the corresponding normally
ordered quantum expectation value. We further simplify the problem by
considering the limit $\gamma_{\mathrm{p}}\gg\gamma_{\mathrm{s}}$, where the
pump variables can be adiabatically eliminated as $\alpha_{0}=\left(
\mathcal{E}_{\mathrm{p}}-\chi\alpha_{+1}\alpha_{-1}\right)  /\gamma
_{\mathrm{p}}$ and $\alpha_{0}^{+}=\left(  \mathcal{E}_{\mathrm{p}}-\chi
\alpha_{+1}^{+}\alpha_{-1}^{+}\right)  /\gamma_{\mathrm{p}}$, arriving at our
model equations:
\end{subequations}
\begin{equation}
\dot{\alpha}_{i}=\gamma_{\mathrm{s}}\left(  -\alpha_{i}+\sigma\alpha_{j}%
^{+}-g^{2}\alpha_{j}^{+}\alpha_{j}\alpha_{i}\right)  +\sqrt{\chi\alpha_{0}}%
\xi_{i}, \label{a}%
\end{equation}
the overdot meaning $\mathrm{d}/\mathrm{d}t$, $i,j=\pm1$\ ($i\neq j$), and the
dimensionless parameters
\begin{equation}
\sigma=\mathcal{E}_{\mathrm{p}}\chi/\gamma_{\mathrm{p}}\gamma_{\mathrm{s}%
},\;g=\chi/\sqrt{\gamma_{\mathrm{p}}\gamma_{\mathrm{s}}}, \label{sigmag}%
\end{equation}
$\sigma^{2}$ being proportional to the external pump power. Another equation
for $\dot{\alpha}_{i}^{+}$ exists that reads $\dot{\alpha}_{i}^{+}=\left(
\dot{\alpha}_{i}\right)  ^{+}$, where the operation "$^{+}$" acts as a
"hermitian-conjugation": complex numbers get complex-conjugated and any
stochastic variable $v$ is transformed according to $v\longleftrightarrow
v^{+}$. In Eqs. (\ref{a}), $\xi_{+1}=\xi_{-1}^{\ast}\equiv\xi$,$\;\xi_{+1}%
^{+}=\left(  \xi_{-1}^{+}\right)  ^{\ast}\equiv\xi^{+}$, and $\left(  \xi
,\xi^{+}\right)  $ are two independent complex noise sources with zero mean
and non zero correlations $\left\langle \xi\left(  t_{1}\right)  \xi^{\ast
}\left(  t_{2}\right)  \right\rangle =\left\langle \xi^{+}\left(
t_{1}\right)  \left[  \xi^{+}\left(  t_{2}\right)  \right]  ^{\ast
}\right\rangle =\delta\left(  t_{1}-t_{2}\right)  $.

\textit{Classical steady emission}.-- The classical DOPO dynamical equations
are obtained by setting $\alpha_{\pm1}^{+}=\alpha_{\pm1}^{\ast}$ and ignoring
noise terms in Eqs. (\ref{a}). Above threshold ($\sigma>1$) the only stable
steady state reads%
\begin{equation}
\bar{\alpha}_{\pm1}=\rho\exp\left(  \mp i\theta\right)  ,\text{ \ }\rho
^{2}=g^{-2}\left(  \sigma-1\right)  \label{classical steady}%
\end{equation}
with $\theta$ an arbitrary phase. The corresponding classical slowly varying
envelope is obtained from Eq. (\ref{As}) after the replacement $\left\{
\hat{a}_{m},\hat{a}_{m}^{\dagger}\right\}  \rightarrow\left\{  \alpha
_{m},\alpha_{m}^{+}\right\}  $ and reads%
\begin{equation}
A_{\mathrm{s}}^{\mathrm{cl}}\left(  \mathbf{r}\right)  =\left(  2\pi
^{-1/2}w^{-2}\rho\right)  r\cos\left(  \phi-\theta\right)  e^{-r^{2}/2w^{2}},
\label{steady}%
\end{equation}
which is a first order H-G mode rotated by $\theta$ with respect to the
transverse $x$ axis, see Fig. 1. The arbitrariness of $\theta$ reflects the
rotational invariance of the problem.

\textit{Quantum fluctuations}.-- The dynamics of quantum fluctuations around
the classical solution are studied by writing $\alpha_{\pm1}=\bar{\alpha}%
_{\pm1}+\delta\alpha_{\pm1}$, and deriving evolution equations for the
fluctuations $\delta\alpha$'s. As $\left\vert \bar{\alpha}_{\pm1}\right\vert
^{2}\gg1$ (these quantities give the classical number of signal photons in
each mode, which are very large above threshold) we assume that $\left\vert
\delta\alpha_{\pm1}\right\vert ,\left\vert \delta\alpha_{\pm1}^{+}\right\vert
\ll\left\vert \bar{\alpha}_{\pm1}\right\vert $ and linearize Eqs. (\ref{a}).
We find it convenient to write the fluctuations as $\delta\alpha_{\pm1}%
=b_{\pm1}e^{\mp i\theta}$, i.e.%
\begin{equation}
\alpha_{\pm1}=\left[  \rho+b_{\pm1}\left(  t\right)  \right]  e^{\mp
i\theta\left(  t\right)  }, \label{bt}%
\end{equation}
with $b_{\pm1}$ (and $b_{\pm1}^{+}$) c-number stochastic variables accounting
for quantum fluctuations. Note that angle $\theta$ is let to vary with time
as, owed to rotational invariance, it is an undamped quantity that is driven
by quantum noise, as we show below. Inserting (\ref{bt}) into (\ref{a}) and
linearizing one easily gets the linearized Langevin equations
\begin{equation}
{\tiny \ }\mathbf{\dot{b}}-2i\rho{\tiny \ }\mathbf{w}_{0}\dot{\theta
}=\mathcal{L}\mathbf{b}+\sqrt{\gamma_{\mathrm{s}}}~\boldsymbol{\xi}\left(
t\right)  , \label{lin}%
\end{equation}
$\mathbf{b}=\operatorname{col}\left(  b_{+1},b_{+1}^{+},b_{-1},b_{-1}%
^{+}\right)  $, $\mathbf{w}_{0}=\frac{1}{2}\operatorname{col}\left(
1,-1,-1,1\right)  $, $\boldsymbol{\xi}=\operatorname{col}\left(  \xi,\xi
^{\ast},\xi^{+},\left[  \xi^{+}\right]  ^{\ast}\right)  $, and the real and
symmetric matrix $\mathcal{L}$ reads
\begin{equation}
\mathcal{L}=-\gamma_{\mathrm{s}}%
\begin{pmatrix}
\sigma & 0 & \sigma-1 & -1\\
0 & \sigma & -1 & \sigma-1\\
\sigma-1 & -1 & \sigma & 0\\
-1 & \sigma-1 & 0 & \sigma
\end{pmatrix}
. \label{L}%
\end{equation}
The eigensystem of $\mathcal{L}$ consists of the Goldstone mode $\mathbf{w}%
_{0}$, whose null eigenvalue reflects the rotational invariance of the system,
of vector $\mathbf{w}_{1}=\frac{1}{2}\operatorname{col}\left(
-1,-1,1,1\right)  $, with eigenvalue $-2\gamma_{\mathrm{s}}$, and of two other
eigenvectors that are unimportant for our present purposes.

\textit{Angular diffusion of the classical field}.-- In order to catch the
dynamics of the pattern orientation angle $\theta$, see Fig. 1, we project the
linear system (\ref{lin}) onto the Goldstone mode $\mathbf{w}_{0}$ and obtain
\cite{eigensystem, nota1}%
\begin{equation}
\dot{\theta}=\sqrt{D_{\theta}}\operatorname{Im}\left(  \xi^{+}-\xi\right)
,\ \ \ D_{\theta}=\frac{\chi^{2}}{4\gamma_{\mathrm{p}}\left(  \sigma-1\right)
}, \label{thetat}%
\end{equation}
with $D_{\theta}$ a diffusion coefficient. Equation (\ref{thetat}) describes a
Wiener process for $\theta$ as already advanced. How fast this diffusion is
can be estimated by evaluating the variance $\left\langle \left[
\theta\left(  t\right)  -\theta\left(  0\right)  \right]  ^{2}\right\rangle
=D_{\theta}t$. Using common values for the system parameters \cite{parameters}
we find $D_{\theta}\sim10^{-6}%
\operatorname{s}%
^{-1}$ for a pump power twice above threshold ($\sigma^{2}=2$). Hence the
rotation of the classical H-G mode will be minute unless the system is
terribly close to threshold. This occurs as the TEM$_{10}$ mode is
macroscopically occupied and hence presents strong inertia to rotations. This
is a most relevant conclusion as a rapid, random rotation of the output
TEM$_{10}$ mode would entail practical difficulties for the homodyne detection
we pass to describe.

\textit{Homodyne detection and squeezing spectrum.--} In order to demonstrate
that the signal field exiting the DOPO exhibits perfect squeezing (within the
linear approach) in the empty H-G mode perpendicular to the macroscopically
emitted one, we consider a balanced homodyne detection experiment, see e.g.
\cite{Gatti}. The noise spectrum $V\left(  \omega\right)  $ of the intensity
difference between the two output ports of the beam splitter, in which the
signal field exiting the DOPO is mixed with a classical, coherent local
oscillator (LO) of frequency $\omega_{0}$, is given by \cite{Gatti,Perez}%
\begin{equation}
V\left(  \omega\right)  =1+2\gamma_{\mathrm{s}}\int_{-\infty}^{+\infty}%
d\tau\left\langle \delta\mathcal{E}\left(  t\right)  \delta\mathcal{E}\left(
t+\tau\right)  \right\rangle e^{-i\omega\tau}, \label{Sout0}%
\end{equation}
where the positive $P$ representation is used to evaluate the stochastic
average, $\delta\mathcal{E}\left(  t\right)  =\mathcal{E}\left(  t\right)
-\left\langle \mathcal{E}\left(  t\right)  \right\rangle $ with%
\begin{equation}
\mathcal{E}\left(  t\right)  =\mathcal{N}^{-1/2}\int d^{2}\mathbf{r}\left(
A_{\mathrm{L}}^{\ast}A_{\mathrm{s}}+A_{\mathrm{L}}A_{\mathrm{s}}^{+}\right)  ,
\end{equation}
$\mathcal{N}=\int d^{2}\mathbf{r}\left\vert A_{\mathrm{L}}\right\vert ^{2}$,
where an argument $\left(  \mathbf{r},t\right)  $ should be understood in all
fields. $A_{\mathrm{L}}\left(  \mathbf{r},t\right)  $ is the LO transverse
envelope, $A_{\mathrm{s}}\left(  \mathbf{r},t\right)  =\sum_{j=\pm1}\alpha
_{j}\left(  t\right)  L_{j}\left(  \mathbf{r}\right)  $ and $A_{\mathrm{s}%
}^{+}\left(  \mathbf{r},t\right)  =\sum_{j=\pm1}\alpha_{j}^{+}\left(
t\right)  L_{j}^{\ast}\left(  \mathbf{r}\right)  $. When the output is
coherent, $V\left(  \omega\right)  =1$ for all $\omega$, defining the shot
noise level. On the other hand $V\left(  \omega_{\ast}\right)  =0$ signals
perfect squeezing (no noise) at $\omega=\omega_{\ast}$ for the quadrature
selected by the LO.

Following the Introduction we choose the LO transverse envelope to be
$A_{\mathrm{L}}\left(  \mathbf{r},t\right)  \propto\frac{\partial}%
{\partial\theta}A_{\mathrm{s}}^{\mathrm{cl}}\left(  \mathbf{r}\right)  $, i.e.%
\begin{equation}
A_{\mathrm{L}}\left(  \mathbf{r},t\right)  =\eta_{\mathrm{L}}e^{i\psi
_{\mathrm{L}}}r\sin\left(  \phi-\theta\left(  t\right)  \right)
e^{-r^{2}/2w^{2}}, \label{ALOF}%
\end{equation}
with $\eta_{\mathrm{L}}$ a real amplitude and $\psi_{\mathrm{L}}$ the LO
phase. This LO is a H-G mode orthogonal, at every time, to the macroscopically
excited one, Eq. (\ref{steady}). Remind that sufficiently above threshold the
diffusion of $\theta$ is negligible and the matching of the LO to the analyzed
mode should not represent any practical problem.

By using LO (\ref{ALOF}) one finds $\delta\mathcal{E}\left(  t\right)
=\sqrt{2}\sin\left(  \psi_{\mathrm{L}}\right)  c_{1}\left(  t\right)  $, with
$c_{1}\left(  t\right)  =\mathbf{w}_{1}\cdot\mathbf{b}\left(  t\right)  $.
Projecting (\ref{lin}) onto $\mathbf{w}_{1}$ gives \cite{eigensystem}
\begin{equation}
\dot{c}_{1}=-2\gamma_{\mathrm{s}}c_{1}-i\sqrt{\gamma_{\mathrm{s}}%
}\operatorname{Im}\left(  \xi^{+}+\xi\right)  , \label{projt1}%
\end{equation}
which allows the evaluation of the squeezing spectrum (\ref{Sout0}). The
result reads%
\begin{equation}
V_{\psi_{\mathrm{L}}}\left(  \omega\right)  =1-\frac{\sin^{2}\left(
\psi_{\mathrm{L}}\right)  }{1+\left(  \omega/2\gamma_{\mathrm{s}}\right)
^{2}}. \label{Sout}%
\end{equation}
We note that for $\psi_{_{\mathrm{L}}}=\frac{\pi}{2}$ Eq. (\ref{Sout})
coincides exactly with the squeezing spectrum of a usual DOPO \textit{at
threshold }\cite{WM}, displaying perfect squeezing ($V=0$) at $\omega=0$: The
phase quadrature of the TEM$_{01}$ mode orthogonal to the TEM$_{10}$ emitted
by the DOPO is perfectly squeezed at zero noise frequency [i.e., perfect
squeezing occurs at the optical frequency $\omega_{0}$, see (\ref{qfield})].
The remarkable difference with usual DOPOs is that the result here reported is
independent of the system parameters (e.g. it is not sensitive to
bifurcations): It is thus a non-critical phenomenon or, in other words,
squeezing needs not be tuned.

Some extra comments are in order: (i) $V_{\psi_{_{\mathrm{L}}}}\left(
\omega\right)  \leq1$ for any LO phase $\psi_{\mathrm{L}}$, i.e. any
quadrature exhibits noise reduction, but $\psi_{\mathrm{L}}=0$ for which
$V_{\psi_{\mathrm{L}}=0}\left(  \omega\right)  =1$; (ii) $V_{\psi_{\mathrm{L}%
}}\left(  \omega\right)  V_{\psi_{\mathrm{L}}+\frac{\pi}{2}}\left(
\omega\right)  <1$, i.e., the two quadratures measured by the LO for
$\psi_{\mathrm{L}}$ and $\psi_{\mathrm{L}}+\frac{\pi}{2}$ are not a Heisenberg
pair, what looks surprising but is understood by the fact that the detected
mode (and thus the LO) is rotating randomly (\cite{arxiv} and the discussion
below after Eq. (\ref{Sout2})); (iii) The detected squeezing is very weakly
dependent on $\psi_{\mathrm{L}}$, e.g. phase uncertainties of $\psi
_{\mathrm{L}}$, around $\psi_{\mathrm{L}}=\frac{\pi}{2}$, as huge as $15%
\operatorname{{{}^\circ}}%
$ lead to $V\left(  \omega=0\right)  \simeq0.067$ (more that $11$dB of noise
reduction), unlike conventional squeezers \cite{9dB,10dB}.

\textit{Influence of imperfections.--} A natural question is whether the above
result is singular in the sense that deviations from perfect rotational
invariance could destroy it. We address this issue by introducing different
cavity losses, $\gamma_{x}$ and $\gamma_{y}$, along two orthogonal transverse
directions. The corresponding Langevin equations read
\begin{equation}
\dot{\alpha}_{i}=\gamma_{\mathrm{s}}\left(  -\alpha_{i}+\kappa\alpha
_{j}+\sigma\alpha_{j}^{+}-g^{2}\alpha_{j}^{+}\alpha_{j}\alpha_{i}\right)
+\sqrt{\gamma_{y}}\xi_{i},
\end{equation}
where noises have been already written in the linear approximation to be used,
$\gamma_{\mathrm{s}}=\frac{\gamma_{y}+\gamma_{x}}{2}$, $\kappa=\frac
{\gamma_{y}-\gamma_{x}}{\gamma_{y}+\gamma_{x}}$ measures how much the
rotational symmetry is externally broken, and $\gamma_{y}>\gamma_{x}$ for
definiteness ($0<\kappa<1$). Above threshold ($\sigma>1-\kappa$) the classical
steady state is $\bar{\alpha}_{+1}=\bar{\alpha}_{-1}=g^{-1}\sqrt{\sigma
+\kappa-1}$, corresponding to a horizontal H-G mode \cite{nota2}, parallel to
the direction of smaller losses. Use of Eq. (\ref{bt}) (now with $\theta=0$)
and considering a vertical H-G mode as a LO one gets%
\begin{equation}
V_{\psi_{\mathrm{L}}=\frac{\pi}{2}}\left(  \omega\right)  =1-\frac{\left(
1-\kappa^{2}\right)  }{1+\left(  1-\kappa\right)  ^{2}\left(  \omega
/2\gamma_{\mathrm{s}}\right)  ^{2}}, \label{Sout2}%
\end{equation}
independently of the pump level, which reduces to (\ref{Sout}) for $\kappa=0$
($\gamma_{y}=\gamma_{x}$). On the other hand one can show easily that
$V_{\psi_{\mathrm{L}}=0}\left(  \omega\right)  V_{\psi_{\mathrm{L}}=\frac{\pi
}{2}}\left(  \omega\right)  =1$, corresponding to a minimum uncertainty state,
as is usual in DOPOs. This confirms that the apparent violation of the
Heisenberg relation in the previous section was related to the detection
scheme (now the LO is kept fixed). Note that $\kappa\neq0$ does not destroy
the squeezing phenomenon we have described before: For example, for
$\kappa=\frac{1}{3}$ ($\gamma_{y}=2\gamma_{x}$, a huge anisotropy indeed),
$V_{\psi_{\mathrm{L}}=\frac{\pi}{2}}\left(  \bar{\omega}=0\right)  \simeq0.11$
($\simeq10$dB of noise reduction), a very large squeezing level. This simple
approach suggests that the phenomenon here presented is very robust.

\textit{Concluding remarks}.-- The spontaneous breaking of the rotational
symmetry around the cavity axis in a type I DOPO above threshold has been
shown to be a means for squeezing light in a non-critical way. The squeezed
mode is a first-order Hermite-Gauss mode orthogonal to the one in which bright
emission occurs. Such mode rotates randomly but very slowly, typically having
a diffusion coefficient $D_{\theta}\sim10^{-6}%
\operatorname{s}%
^{-1}$. Outstandingly the squeezing level, perfect in the linear approach at
zero noise frequency, is independent of the system's distance from threshold.
This result is robust versus deviations from perfect rotational symmetry, e.g.
due to a cavity anisotropy. The fact that the squeezed mode is a first-order
Hermite-Gauss mode can be of utility for precision measurements \cite{Fabre}.

This work has been supported by the Spanish Ministerio de Educaci\'{o}n y
Ciencia and the European Union FEDER through Project FIS2005-07931-C03-01.
C.N.-B. is grant holder of the Programa FPU del Ministerio de Educaci\'{o}n y
Ciencia (Spain). Fruitful discussions with N. Treps and H. Bachor are acknowledged.


\begin{thebibliography}{99}
\bibitem {chi}Assuming the crystal length $l$ to be short enough as compared
to the cavity Rayleigh length one has $\chi=\left(  3\pi\chi^{\left(
2\right)  }l/w\right)  \sqrt{\hbar/\varepsilon_{0}}\left(  c/n\lambda
L\right)  ^{3/2}$, with $\chi^{\left(  2\right)  }$ the relevant second order
susceptibility and $\lambda$ the pump wavelength.
\bibitem {nota1}We can choose null the projection of $\mathbf{b}\left(
t\right)  $ onto the Goldstone mode as it merely redefines the angle $\theta$.
\bibitem {parameters}We consider a symmetric cavity of length $L=0.1\mathrm{%
\operatorname{m}%
}$ and mirrors' radii $R=1%
\operatorname{m}%
$, pumped at $\lambda=400%
\operatorname{nm}%
$, $\chi^{\left(  2\right)  }=2\frac{%
\operatorname{pm}%
}{\mathrm{%
\operatorname{V}%
}}$, $n=2.5$, $l=1%
\operatorname{mm}%
$, and $\gamma_{\mathrm{p}}=0.16%
\operatorname{ns}%
^{-1}$ ($10\%$ transmision through the pump output coupler).
\end{thebibliography}
\end{document}